\documentclass[11pt]{article}
\usepackage{moriond,epsfig}
\usepackage{rotating,epsf}

\def\Journal#1#2#3#4{{#1} {\bf #2}, #3 (#4)}

\def\EPJ{{\em Eur. Phys. J.}  C}

\def\PLB{{\em Phys. Lett.}  B}
\def\PRL{\em Phys. Rev. Lett.}
\def\PRD{{\em Phys. Rev.} D}

\def\ko{K^0}
\def\kb{\bar{K^0}}

\def\be{\begin{equation}}
\def\ee{\end{equation}}
\def\bea{\begin{eqnarray}}
\def\eea{\end{eqnarray}}

\def\epsoveps{Re(\epsilon' / \epsilon)} 
\def\taus{\tau_s}
\def\delm{\Delta m}
\def\phipm{\phi_{+-}}
\def\delphi{\Delta \phi}
\def\ktoeemm{K_L \rightarrow e^+ e^- \mu^+ \mu^-}
\def\ktomm{K_L \rightarrow \mu^+ \mu^-}
\def\ktommg{K_L \rightarrow \mu^+ \mu^- \gamma}
\def\ktoeeg{K_L \rightarrow e^+ e^- \gamma}

\def\ktotwopi{K_{S,L} \rightarrow \pi \pi}
\def\ktopipi{K \rightarrow \pi \pi}
\def\ktoplmn{K_{S,L} \rightarrow \pi^+ \pi^-}
\def\ktozz{K_{S,L} \rightarrow \pi^0 \pi^0}
\def\kgstarg{K\gamma^* \gamma}
\def\kgstargstar{K\gamma^* \gamma^*}
\def\kztokzbar{\ko \rightarrow \kb}
\def\kzbartokz{\kb \rightarrow \ko}
\def\kltoplmn{K_L \rightarrow \pi^+ \pi^-}
\def\kltozz{K_L \rightarrow \pi^0 \pi^0}
\def\kltozzz{K_L \rightarrow \pi^0 \pi^0 \pi^0}
\def\kstoplmn{K_S \rightarrow \pi^+ \pi^-}
\def\kstozz{K_S \rightarrow \pi^0 \pi^0}
\def\ktollg{K_L \rightarrow \ell^+ \ell^- \gamma}
\def\ktollll{K_L \rightarrow \ell^+ \ell^- \ell^+ \ell^-}
\def\kethree{K_L \rightarrow \pi e \nu}
\def\kteveps{(20.71 \pm 1.48 (stat) \pm 2.39 (syst))~\times~10^{-4}}
\def\ktevepsf{(20.7 \pm 2.8) \times 10^{-4}}
\def\ktevphipm{(44.12 \pm 0.72 (stat) \pm 1.20 (syst))^\circ}
\def\ktevdelphi{(+0.39 \pm 0.22 (stat) \pm 0.45 (syst))^\circ}
\def\ktevdelm{(5288 \pm 23 (stat))~\times~10^6 \hbar~s^{-1}}
\def\ktevtaus{(89.58 \pm 0.08 (stat))~\times~10^{-12}~s}

\begin{document}
\vspace*{4cm}
\title{KTeV Results: $\epsoveps$ and Rare Decay Results}

\author{ J. Whitmore, for the KTeV Collaboration }

\address{Fermi National Acceleratory Laboratory, Box 500, MS-205,\\
Batavia, IL  60510 USA}

\maketitle\abstracts{
The KTeV physics program encompasses 
many goals including a precision measurement of the
direct CP violation parameter $\epsoveps$ in $\ktotwopi$ decays, 
and studies of rare neutral kaon decays. 
The KTeV detector collected data
during the Fermilab fixed-targed runs of 1996-97 and 1999.  
This article focuses on the precision measurement of  
the direct CP violation parameter $\epsoveps$ using the 1996-97 
data set.  In addition, measurements of the neutral kaon parameters 
$\taus$, $\delm$, $\phipm$, $\delphi$ from that data set and 
a new measurement of the branching fraction of $\ktoeemm$ 
from the 1997 and 1999 data also are presented.
}

\section{Introduction}\label{sec:intro}
CP violation in the $\ktotwopi$ system comes mainly from a small 
($\sim 10^{-3}$) asymmetry between the transition rates of 
$\kztokzbar$ and $\kzbartokz$, which is
referred to as indirect CP violation.  Direct CP violation is 3 orders of 
magnitude smaller and is CP violation in the decay amplitude itself.  
Experimentally, direct CP violation
can be determined by comparing decay rates for $\ktoplmn$ and $\ktozz$
in the following double ratio
\be
{{\Gamma(\kltoplmn)/\Gamma(\kstoplmn)} \over 
{\Gamma(\kltozz)/\Gamma(\kstozz)}} = 
\left| {\eta_{+-} \over \eta_{00}} \right|^2
\approx 1 + 6 \epsoveps,
\label{eq:epsoveps}
\ee
where $\epsilon$ refers to the parameter that 
governs indirect CP violation, and $\epsilon'$ is the parameter that
indicates the strength of direct CP violation.
The double ratio allows for many systematic uncertainties 
in the measurement to cancel.
The KTeV experiment was designed to make a precise determination
of $\epsoveps$.  
KTeV can also measure the $K_S$ lifetime ($\taus$), $K_L - K_S$ mass 
difference ($\delm$), the phase of $\eta_{+-}$ ($\phipm$), and 
the phase difference between $\eta_{00}$ and $\eta_{+-}$ ($\delphi$).

In addition to being optimized for the detection of $\ktotwopi$ decays, 
the experiment is ideally suited for the detection of rare
kaon decays such as $\ktollg$ and $\ktollll$, where $\ell$ is either an 
electron or a muon.  These decays are of interest because they can provide 
insight into the long distance contributions to the decay $\ktomm$, which is 
sensitive to the CKM parameter $V_{td}$.

\begin{figure}[tbh]
\begin{centering}
\epsfig{figure=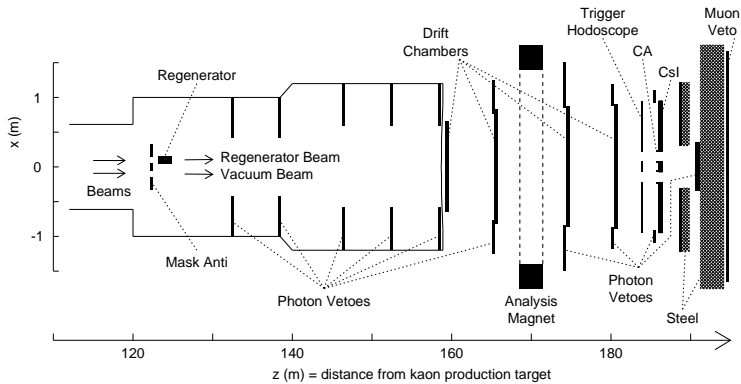,height=2in}
\caption{A plan view of the KTeV Detector configured for $\epsoveps$ 
data taking.
\label{fig:det}}
\end{centering}
\end{figure}
The KTeV detector is shown in Fig.~\ref{fig:det}.
The experiment features a 65m vacuum decay region followed by a large aperture 
charged spectrometer, consisting of 4 planar drift chambers, two on either
side of a dipole magnet. 
Downstream is a 2m-by-2m calorimeter comprised of 3100 pure CsI crystals.
Behind the calorimeter is 10cm of Pb, 5m of steel and a set of 
muon identification counters.
Photon veto detectors surround the vacuum decay region, the perimeter
of the drift chambers, and the CsI.
During $\epsoveps$ run periods, an additional upstream 
photon veto detector (Mask-Anti) and an active regenerator are used; 
the regenerator alternates beam positions once per minute
to cancel possible left/right detector and beam asymmetries.  
A set of 8 transition radiation detectors (TRD) are placed 
in front of the CsI during rare decay running for improved $e/\pi$ 
identification.  The charged spectrometer achieves a hit resolution of 
better than $100 \mu$m, while the CsI calorimeter gets better 
than $1\%$ energy resolution over the energy range of interest.

The KTeV experiment took data for the $\epsoveps$ measurement
and for rare decays during 1996-97 and 1999.  Between the 
two running periods, the detector was upgraded to improve its reliability and 
live time. 
During the $\epsoveps$ data collecting, the transverse momentum 
kick of the analysis magnet was 410 MeV/c.  The momentum kick for rare decay
run periods was reduced from 205 MeV/c in 1997 to 150 MeV/c in 1999 to
increase the acceptance for higher multiplicity modes. 
During 1996-97 $\epsoveps$ data taking, 
KTeV collected $3.3~\times~10^6$ $\kltozz$ events, 
which is the statistics limiting mode for the measurement.
This paper presents $\epsoveps$ results from 
the 1996-97 data set~\cite{eps}.
For rare decay running, there were $2.7~\times~10^{11}$ and 
$3.6~\times~10^{11}$ 
$K_L$ decays during 1997 and 1999, respectively.
In the case of the $\epsoveps$ analysis, where published results are now 
available, numbers have been updated from those presented 
at the conference.

\section{Measurements of $\epsoveps$ and the Kaon Parameters}
\label{sec:epsoveps}

The $\epsoveps$ measurement requires a source of $K_L$ and $K_S$ decays.
The regenerator produces a coherent $|K_L> + \rho |K_S>$ state; 
in KTeV, the regeneration amplitude, $\rho$, is 
$\sim 0.03$ (at 70 GeV/c) and is momentum dependent. 
The measured quantities for the double ratio 
in Eq.~\ref{eq:epsoveps} are extracted with a fitting program, where 
the $K_S-K_L$ interference in the regenerator beam is accounted for.
The fitting for $\epsoveps$ is performed in 10 GeV/c kaon momentum bins
and a single, integrated z bin from 110m to 158m.  
The $p$ binning reduces the sensitivity to 
the momentum dependence of the detector acceptance and allows
the momentum dependence of the regeneration amplitude, $\rho$, 
to be accounted for in the fit.  

Because the $K_L$ and $K_S$ events have very different lifetimes and 
produce distinctly different decay distributions in the detector, 
a key to making the $\epsoveps$ measurement is understanding 
the acceptance.
A detailed Monte Carlo simulation
has been developed to understand the acceptance 
correction to the measurement.  The acceptance correction
shifts $\epsoveps$ by $\sim~85~\times~10^{-4}$;
$\sim~85\%$ of this correction is the result of 
detector geometry that is known precisely from surveys and 
from measurements in the data, while the remaining correction depends
on details of the detector response and resolutions.

\begin{figure}
\begin{centering}
\psfig{figure=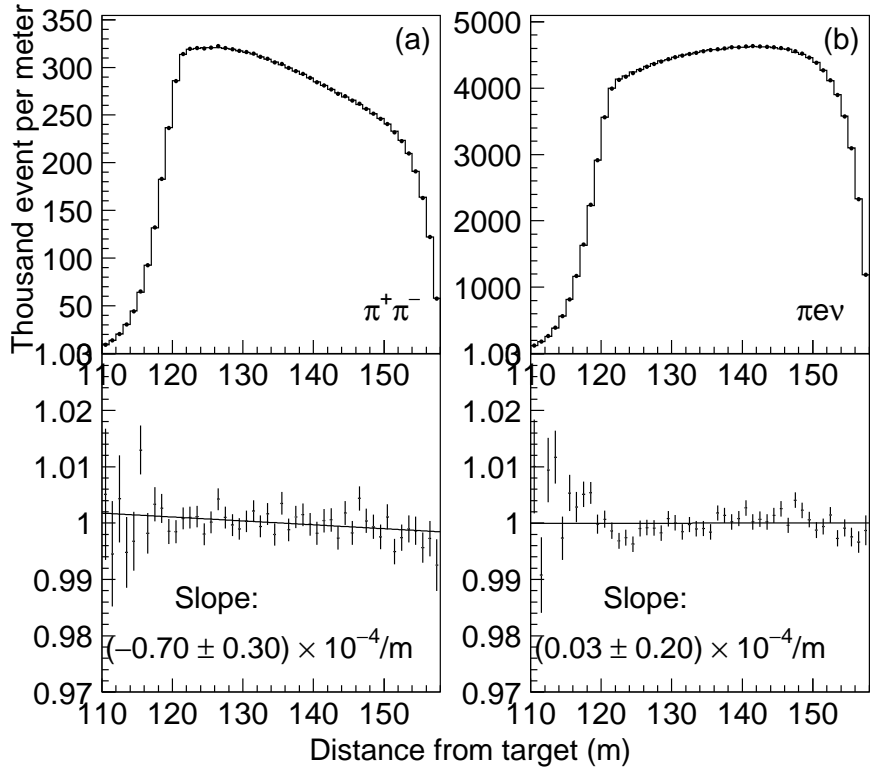,height=4.25in}
\caption{Comparison of the vacuum beam z distributions for 
data(dots) and MC (histogram) for $\kltoplmn$ and $\kethree$ decays.  
Data/MC ratios are shown with the results of the linear fits.
The $\kltoplmn$ fit shows a $2.3\sigma$ slope, while the high statistics
$\kethree$ sample is consistent with having zero slope.
\label{fig:zvtx}}
\end{centering}
\end{figure}

An important test of the acceptance is to compare distributions of 
the reconstructed Z-vertex positions in data and Monte Carlo.  
The charged and neutral $\ktopipi$ samples are studied as well as the
high statistics $\kethree$ and $\kltozzz$ modes.
Fig~\ref{fig:zvtx} shows the data/Monte Carlo 
ratio for $\kltoplmn$ and $\kethree$ decays along with the results
to the linear fit.
A $2.3\sigma$ ``z-slope'' is measured in the $\kltoplmn$ events while
the $\kethree$ mode is consistent with zero slope.  
This translates to a 
$0.79~\times~10^{-4}$ uncertainty in $\epsoveps$ and 
is the largest contribution
to the systematic uncertainty from charged mode decays.
No significant ``z-slopes'' are measured for the neutral mode samples, 
resulting in a systematic uncertainty in $\epsoveps$ due to the 
z-dependence of the acceptance of $0.39~\times~10^{-4}$.

\begin{figure}
\begin{centering}
\psfig{figure=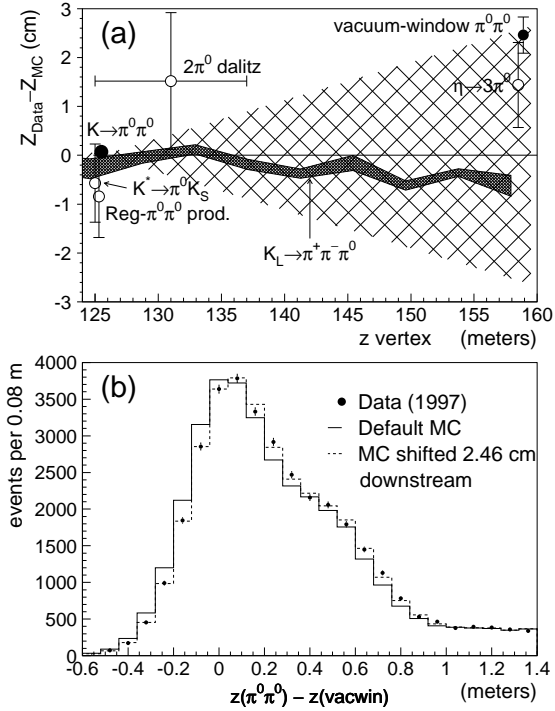,height=3.75in}
\caption{(a) The difference between the reconstructed 
data and MC neutral vertex at different locations along the decay region.
The hatched region shows the range of discrepancies covered by the systematic
error.
(b) The reconstructed $\pi^0\pi^0$ vertex distribution at the  
the vacuum window for data after matching the regenerator edge,
for the default MC, and after shifting the MC 2.46cm downstream.
\label{fig:escale}}
\end{centering}
\end{figure}

The largest contribution to the systematic uncertainty on $\epsoveps$ 
comes from the understanding of the neutral energy scale.
The neutral mode energy calibration is determined using electrons from 
$\kethree$ decays.
The final energy scale adjustment is achieved by 
matching the downstream regenerator edge in data and MC
for $\kstozz$ decays.
The regenerator edge shift is determined 
in $10~GeV$ kaon energy bins.  On average, the data edge is 
shifted upstream by 5~cm relative to the MC,  
corresponding to a ~0.1\% correction on the
energy scale. 
A series of other decay modes are used to check the energy scale at other
z positions (See Fig~\ref{fig:escale} (a)).
The uncertainty in $\epsoveps$ due to the neutral energy scale 
is determined by 
introducing an energy scale distortion to data such that data and MC z-vertex
distributions are made to match at both the downstream regenerator 
(fixed by construction) and the vacuum-window edges.
This distortion introduces a shift to $\epsoveps$ 
of $-1.08~\times~10^{-4}$ and $-1.37~\times~10^{-4}$ to 1996 and 1997 data 
sets, respectively, and results in a systematic uncertainty of 
$1.27~\times~10^{-4}$.  
Additional contributions from energy non-linearities and reconstructed 
cluster position discrepancies lead to an overall systematic uncertainty
on $\epsoveps$ from neutral energy reconstruction of $1.47~\times~10^{-4}$.

Table~\ref{tab:syst} summarizes the dominant contributions to the uncertainty 
of $\epsoveps$.
\begin{table}[t]
\caption{Dominant systematic uncertainties of $\epsoveps$.
Neutral mode background from 
regenerator-scatters refers to a $\ktopipi$ decay from a kaon that has 
scattered in the regenerator, which is described in the MC using a 
function that is fit to the acceptance-corrected, background-subtracted
$K \rightarrow \pi^+ \pi^-$ decays.
The contribution to the uncertainty 
from the Level 3 online filtering for charged mode decays
is determined by comparing the shift in vacuum-to-regenerator ratio 
after applying Level 3 requirements on data and MC events.  
\label{tab:syst}}
\vspace{0.4cm}
\begin{center}
\begin{tabular}{|c|c|}
\hline
Source& Contribution to uncertainty \\
      &  on $\epsoveps$ $(\times~10^{-4})$ \\ \hline
Neutral Energy Reconstruction& 1.47 \\ 
Neutral background from regenerator-scatters & 1.07 \\
Charged mode acceptance (data/MC z-slope) & 0.79 \\
Charged Level 3 online filter & 0.58 \\ \hline
\end{tabular}
\end{center}
\end{table}
The combined result for $\epsoveps$ for the 1996-97 data sets is 
\be
\kteveps.
\ee
A crosscheck of the standard $\epsoveps$ analysis was performed 
using a reweighting method that does not depend on a Monte Carlo 
acceptance correction.  
\footnote{This technique is similar to the NA48 experimental 
method~\cite{na48}.}
The difference between the reweighting and standard $\epsoveps$ results is
$\Delta[\epsoveps] = 
+1.5 \pm 2.1 (stat) \pm 3.3 (syst)]~\times~10^{-4}$,  
where the errors presented are the uncorrelated uncertainties.  
There is good agreement between the two analyses, although the statistical
uncertainty of the reweighting analysis is a factor of 1.7 larger than the
standard analysis because vacuum beam events are ``lost'' in the reweighting.
In addition, the uncorrelated systematic uncertainties between the two 
methods are somewhat large, dominated by a sensitivity to low energy 
clusters in the reweighting analysis.

The regenerator beam decay distribution allows for measurements of 
the kaon parameters $\delm$, $\taus$, $\delphi$ and $\phipm$.
The parameters are measured by fitting the shape of the z decay 
distribution in the regenerator beam 
in 2m z-bins from 124-158m using the flux determined from the vacuum beam.  
The results are listed in Table~\ref{tab:kparam}.
\begin{table}[t]
\caption{Results from fitting the regenerator beam z distribution for 
the kaon parameters.
\label{tab:kparam}}
\vspace{0.4cm}
\begin{center}
\begin{tabular}{|c|c|}
\hline
Measurement& Results \\ \hline 
$\taus$ & $\ktevtaus$ \\ 
$\delm$ & $\ktevdelm$ \\
$\phipm$ & $\ktevphipm$ \\
$\delphi$ & $\ktevdelphi$ \\ \hline
\end{tabular}
\end{center}
\end{table}

\section{The Decay $\ktoeemm$}\label{sec:ktoeemm}
\begin{figure}
\begin{centering}
\psfig{figure=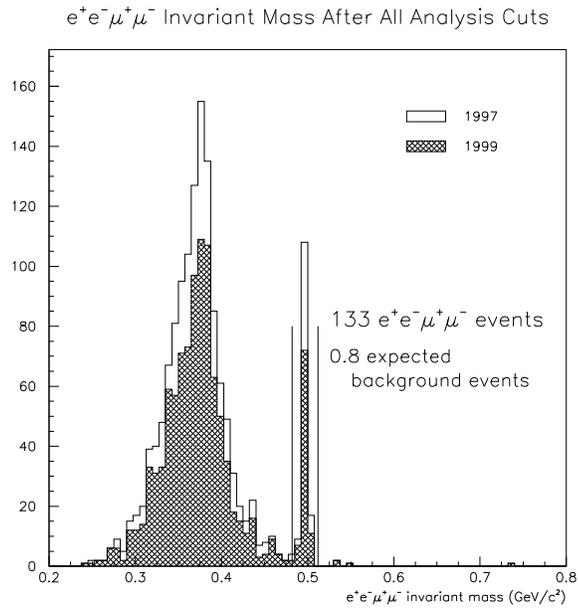,height=3.5in}
\caption{Invariant mass distribution for $\ktoeemm$ candidates for 
the 1997 and 1999 data sets.
\label{fig:masseemm}}
\end{centering}
\end{figure}

The decay $\ktoeemm$ provides the cleanest method for studying 
the $\kgstargstar$ vertex, which is useful for determining the long 
distance contributions to the $\ktomm$ decay.  But because 
of its extremely small branching ratio~\cite{ping,amitl}, it is 
difficult to collect a sample large enough to probe the $\kgstargstar$
form factor.  The combined KTeV 1997 and 1999 rare decay running periods  
yielded 132 $\ktoeemm$ events (see Fig.~\ref{fig:masseemm}), 
corresponding to a 
preliminary branching ratio measurement~\cite{jhamm} of 
BR($\ktoeemm$) = $(2.68 \pm 0.23 \pm 0.12) \times~10^{-9}$.  

\begin{figure}
\begin{centering}
\psfig{figure=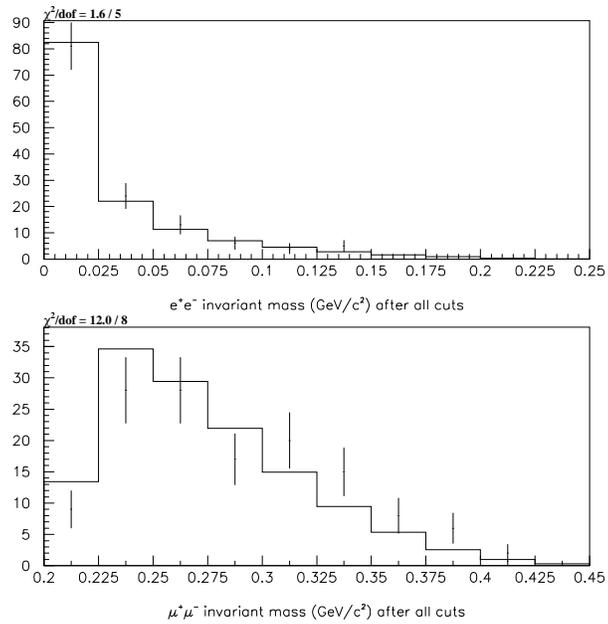,height=3.5in}
\caption{The $e^+e^-$ and $\mu^+\mu^-$ invariant mass distributions from
$\ktoeemm$ decays.  The data indicate a non-trivial form factor.
\label{fig:eemmff}}
\end{centering}
\end{figure}

Although the $\ktoeemm$ sample contains enough events to study 
the $\kgstargstar$ form factor by fitting the $e^+e^-$ and $\mu^+\mu^-$ 
mass distributions (See Fig.~\ref{fig:eemmff}), the resulting uncertainties 
are expected to be larger than with other 
modes~\cite{mmg,eeee}.
The $\ktoeemm$ decays are being analyzed to determine the 
form factors $\alpha_K*$ from the Bergstrom, Masso and Singer (BMS) model and 
$\alpha$ from the D'Ambrosio, Isidori, and Pertoles (DIP) 
model~\cite{bms,dip}.
Because of the limited statistics, it is doubtful that
a measurement of the DIP parameter $\beta$ will be possible.

\section{Conclusions}\label{sec:conclusion}
\begin{figure}
\begin{centering}
\psfig{figure=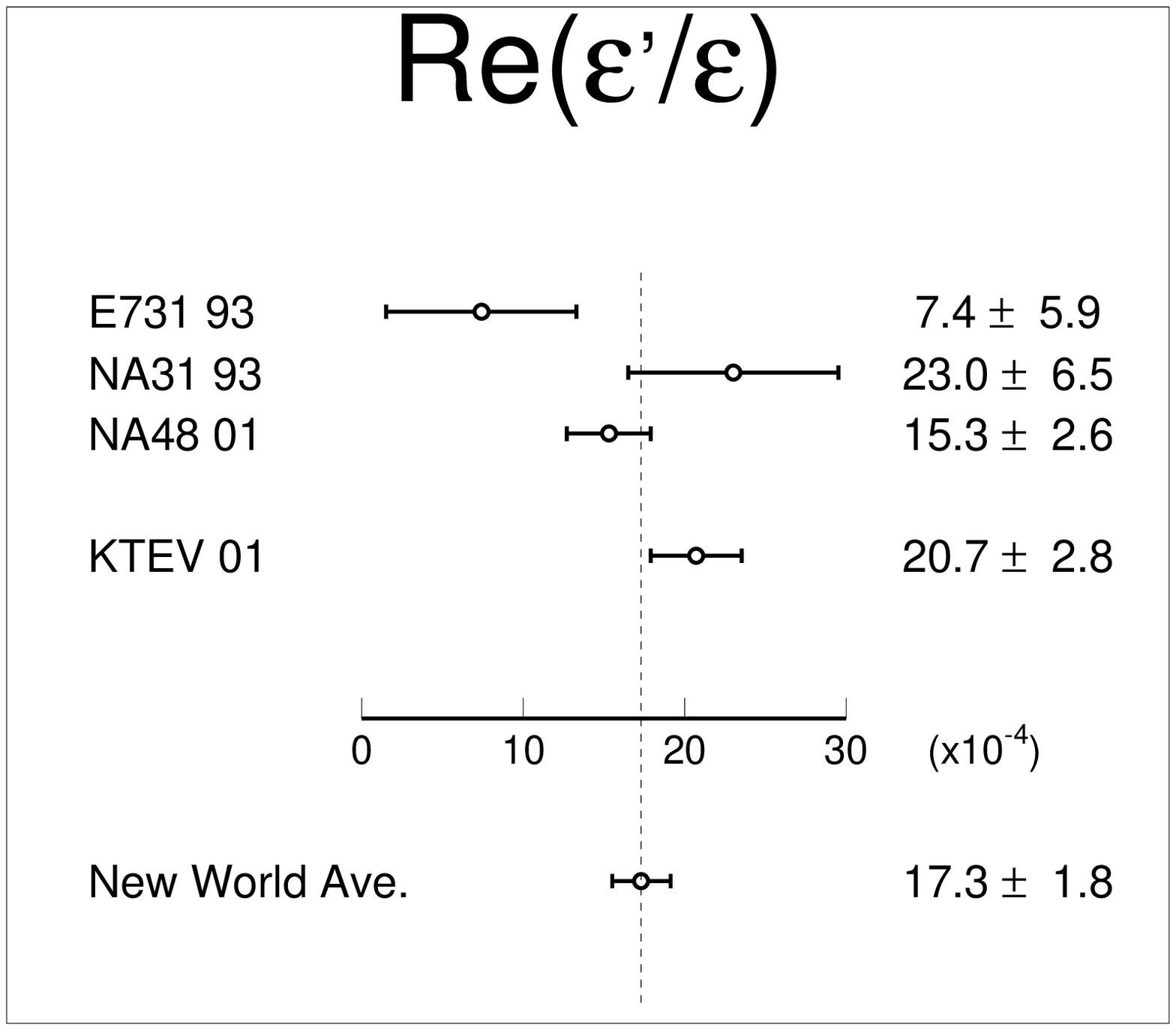,height=2.5in}
\caption{The status of the current $\epsoveps$ measurements.
\label{fig:eps}}
\end{centering}
\end{figure}

KTeV has made a precision measurement of the direct CP violation
parameter $\epsoveps$ of $\ktevepsf$. 
The status of the current measurements~\cite{eps,na48,e731,na31} is summarized 
in Fig~\ref{fig:eps}.
The current precision of the world average of $\epsoveps$ measurements is
better than 10\%.  These measurements firmly establish direct CP violation 
at the $10\sigma$ level.  
KTeV has also made precision measurements of the kaon parameters 
$\delm$, $\taus$, $\delphi$ and $\phipm$ using their 1996-97 data set.
Analysis of the KTeV 1999 data set is ongoing.  
Inclusion of the 1999 data set will double the statistics of the 1996-97 
result.  Improvements in the neutral energy reconstruction are expected 
to reduce the systematic error.  

In addition to the $\epsoveps$ program, KTeV has made a significant
contribution to the understanding of rare kaon decays.  Decays that were 
once considered ``rare'' , such as $\ktommg$ with a branching ratio of 
$10^{-7}$, are now represented by data samples of 
$\sim10K$ events.  In the case
of the decay $\ktoeeg$, the combined 1997 and 1999 data sets have yielded
$\sim127K$ events, allowing for detailed studies of the $\kgstarg$ 
form factor.  
With statistically large samples such as these, 
precision rare decay measurements are achievable.
 
\section*{Acknowledgments}
The author would like to thank 
Elliott Cheu and Ed Blucher for their contributions 
to this paper.

\end{document}